# An edge detection-based deep learning approach for tear meniscus height measurement

Kesheng Wang[a,#], Kunhui Xu[b,#], Xiaoyu Chen[b], Chunlei He[a], Jianfeng Zhang[a], Dexing Kong[a], Qi Dai[a,b,*] and Shoujun Huang[a,c,*]

[a]College of Mathematical Medicine, Zhejiang Normal University, Jinhua 321004, P. R. China

[b]National Clinical Research Center for Ocular Diseases, Eye Hospital, Wenzhou Medical University, Wenzhou 325027, P. R. China

[c]Puyang Big Data and Artificial Intelligence Institute, Puyang 457000, P. R. China

*Corresponding author: Qi Dai, dq@mail.eye.ac.cn; Shoujun Huang, sjhuang@zjnu.edu.cn

[#]These authors share first authorship

**Abstract**

Automatic measurements of tear meniscus height (TMH) have been achieved by using deep learning techniques; however, annotation is significantly influenced by subjective factors and is both time-consuming and labor-intensive. In this paper, we introduce an automatic TMH measurement technique based on edge detection-assisted annotation within a deep learning framework. This method generates mask labels less affected by subjective factors with enhanced efficiency compared to previous annotation approaches. For improved segmentation of the pupil and tear meniscus areas, the convolutional neural network Inceptionv3 was first implemented as an image quality assessment model, effectively identifying higher-quality images with an accuracy of 98.224%. Subsequently, by using the generated labels, various algorithms, including Unet, ResUnet, Deeplabv3+FcnResnet101, Deeplabv3+FcnResnet50, FcnResnet50, and FcnResnet101 were trained, with Unet demonstrating the best performance. Finally, Unet was used for automatic pupil and tear meniscus segmentation to locate the center of the pupil and calculate TMH, respectively. An evaluation of the mask quality predicted by Unet indicated a Mean Intersection over Union of 0.9362, a recall of 0.9261, a precision of 0.9423, and an F1-Score of 0.9326. Additionally, the TMH predicted by the model was assessed, with the fitting curve represented as $y = 0.982x - 0.862$, an overall correlation coefficient of $r^2 = 0.961$, and an accuracy of 94.80% (237/250). In summary, the algorithm can automatically screen images based on their quality, segment the pupil and tear meniscus areas, and automatically measure TMH. Measurement results using the AI algorithm demonstrate a high level of consistency with manual measurements, offering significant support to clinical doctors in diagnosing dry eye disease.

# 1. Introduction

Dry eye disease (DED) is a multifactorial condition[1]. The Tear Film and Ocular Surface Society (TFOS) Dry Eye Workshop (DEWS) II identified the central mechanism of dry eye disease as evaporative water loss leading to hyperosmolar tissue damage. Their research indicates that DED, either directly or through the induction of inflammation, can result in the loss of epithelial and goblet cells. This loss leads to reduced ocular surface wettability, causing premature tear film rupture, which in turn increases osmotic pressure and creates a vicious cycle [2]. DED is one of the most prevalent diseases in ophthalmology. A global prevalence study by the TFOS DEWS II subcommittee revealed that DED affects 5–50% of the population [3]. As the population ages and the use of electronic devices becomes increasingly prevalent, the incidence of DED is steadily increasing. The condition, which varies in severity, causes a range of discomfort in affected individuals. Therefore, efficient, accurate, and convenient diagnosis of DED is of significant importance.

Evaluation of the tear meniscus is currently a highly effective approach in the diagnosis of DED. Tear meniscus height (TMH), a crucial parameter for evaluating the tear meniscus, has been extensively researched in recent years to explore its relationship with DED. In 2007, Uchida et al. utilized a tear interference device (Tearscope plus, Keeler, Windsor, United Kingdom) to compare the TMH in normal individuals and patients with DED. They observed that TMH in patients with DED is generally significantly lower than that in normal individuals [4]. In 2010, Yuan et al. used optical coherence tomography (OCT) to measure tear meniscus dynamics in DED patients with aqueous tear deficiency and concluded that the TMH in patients with DED is lower than that in normal individuals under both normal and delayed blinking conditions [5]. In 2016, Tian et al.

investigated the repeatability and reproducibility of TMH measurements using Keratograph 5M. Their study confirmed that TMH measurement with Keratograph5M offers acceptable repeatability and reproducibility [6]. The Keratograph5M is now a commonly used clinical device to acquire image data for DED screening.

Numerous relevant studies have been conducted regarding the implementation of automated TMH measurement algorithms. In 2019, Stegmann et al. used a custom OCT system and a threshold-based segmentation algorithm (TBSA) to examine the lower tear meniscus [7]. Based on this, in 2020, Stegmann et al. developed a deep learning model using TBSA-segmented tear meniscus images [8]. Additionally, in 2019, Arita et al. devised and assessed a method using the Kowa DR-1a tear interferometer for quantitative TMH measurement. However, this method requires point selection by the operator before calculation [9]. In 2019, Yang et al. introduced a novel automatic image recognition software that uses a threshold-based algorithm for TMH measurement, although image data collection in this method is somewhat invasive [10]. Moreover, in 2021, Deng et al. proposed a fully convolutional neural network-based approach for automatic segmentation of the tear meniscus area and TMH calculation. However, this method employed polynomial functions to delineate the overall upper and lower boundaries of the tear meniscus, introducing significant errors [11]. In 2023, Wan et al. designed an algorithm for tear meniscus area segmentation using the DeepLabv3 structure, enhanced with elements of the ResNet50, Google-Net, and FCN network structures. This study, however, was limited by a small data volume and the relative complexity of the DeepLabv3 model structure [12]. The two aforementioned deep learning techniques trained the pupil segmentation model and the tear meniscus segmentation model separately. When using images captured by Keratograph5M, the TMH calculation model heavily relies on expert-level, pixel-level annotations; the precision of these annotations is crucial for the effectiveness of training the neural network model. Generating such detailed annotations demands considerable time and effort from experts and presents a formidable challenge.

The primary objective of this study is to facilitate the generation of mask labels using a method that is more efficient, less labor-intensive, and robust. This approach aims to enable the simultaneous segmentation of the tear meniscus and pupil areas using a single model, along with the calculation of TMH. Additionally, we evaluated the effectiveness and practicality of the model in assisting physicians with the diagnosis and treatment processes by analyzing the TMH measured with our algorithm.

# 2. Methods

## 2.1 Statistical analysis methods

Statistical analysis was conducted using the SciPy stats package (version 1.9.1) and Numpy (version 1.21.5) in Python. To facilitate the comparison between the measured TMH and the ground-truth, all data analyses involving lengths in this study were conducted in pixels. In our experimental data, one pixel equals approximately 0.011575 mm. Prior to model development, we analyzed the data using the Intraclass Correlation Coefficient (ICC) [13], Pearson correlation analysis, and Spearman correlation analysis [14] to ensure the value and reliability of the data for model development. For evaluating model performance, we employed Mean Intersection over Union (MIoU), recall, precision, and F1-Score to assess the accuracy of the model's predicted mask. Additionally, Accuracy (ACC), linear fitting, and Bland-Altman plots [15] were used to evaluate and illustrate the accuracy of the TMH measurements obtained by the model.

(1) Interclass Correlation Coefficient (ICC)[13]

This paper used a 2-way fixed effects model to calculate the absolute agreement (which indicates the similarity of scores for the same group of participants) for a single rater and the mean of k raters. The formula is as follows:

$$c_1 = \frac{MSR - MSE}{MSR + (k-1)MSE + \frac{k}{n}(MSC - MSE)},$$

$$c_2 = \frac{MSR - MSE}{MSR + \frac{(MSC - MSE)}{n}},$$

Where $MSR$ represents the row mean square, $MSE$ the error mean square, $MSC$ the column mean square, $n$ the number of test samples, and $k$ the number of raters.

(2) Pearson correlation test

The Pearson correlation test is a statistical method to measure the linear correlation between two variables. The formula is as follows:

$$\rho(X,Y) = \frac{Cov(X,Y)}{\sigma_X \cdot \sigma_Y},$$

Where $\rho(X,Y)$ represents the Pearson correlation coefficient between $X$ and $Y$, $Cov(X,Y)$ represents the covariance between $X$ and $Y$, $\sigma_X$ represents the standard deviation of $X$, and $\sigma_Y$ represents the standard deviation of $Y$.

**(3) Spearman correlation test[14]**

The Spearman correlation test is a non-parametric statistical method based on the rank of sorted data to measure the correlation between two variables. It is more robust to non-linear relationships or outliers. The formula is as follows:

$$\rho = 1 - \frac{6\sum d_i^2}{n(n^2-1)},$$

Where $\rho$ is the correlation coefficient, $d$ is the difference in ranks for each pair, and $n$ is the sample size.

**(4) Accuracy (ACC) of TMH**

The accuracy of the measured TMH was determined by comparing it with the ground-truth. It was calculated as the ratio of the number of accurate values to the total number of measured images. A height difference within $\pm 3$ pixels ($\pm 0.0347$ mm) was considered accurate.

## 2.2 Preprocessing and annotation methods

This study involved a total of 500 patients and 1,000 ocular surface images. All images were obtained from the Eye Hospital, Wenzhou Medical University from July 2020 to July 2023. This study was approved by the Ethics Review Board (IRB) of Eye Hospital, Wenzhou Medical University (IRB approval number: H2023-045-K-42), and all procedures were in strict compliance with the principles of the Declaration of Helsinki. All ocular surface images were captured using a Keratograph5M (K5M). During the image capturing, patients were guided by the skilled technician to position their chin 100 mm in front of the camera. The data were saved in PNG format with a resolution of 1024×1360 pixels in RGB channels. The images were renamed using a fixed coding system to ensure personal privacy. Experimental hardware configuration for training and testing the models: 20 Intel(R) Xeon(R) W-2255 CPUs @ 3.70GHz and NVIDIA RTXA4000. Experimental software configuration: Ubuntu 22.04.2 LTS, PyCharm 2023.1.4 Professional Edition, and Python 3.8.17.

In this study, the upper and lower edges of the stripe-shaped tear film near the lower eyelid margin were defined as the upper and lower boundaries of the tear meniscus, respectively. The center point within the smallest ring of the Placido ring was defined as the center of the pupil area. An initial screening of 1,000 images was conducted, resulting in the exclusion of 166 poor-quality images. From the remaining 834 images, 200 were randomly selected and independently annotated by three experts labeled A, B, and C. Each expert generated masks for the tear meniscus area and the pupil center and measured the horizontal coordinates of the TMH and the pupil center. This annotation process was repeated three times, with no time constraint for completion and intervals between sessions, to minimize the influence of previous annotations. Experts A and B used a pixel-by-pixel approach to generate the tear meniscus area mask and submitted their work for review by a more senior expert. Expert C adopted the edge detection-based method proposed in this study to generate the tear meniscus mask, with the annotation process illustrated in Fig. 1.We will compare the annotations of three experts to demonstrate that our method is helpful to produce better annotations.

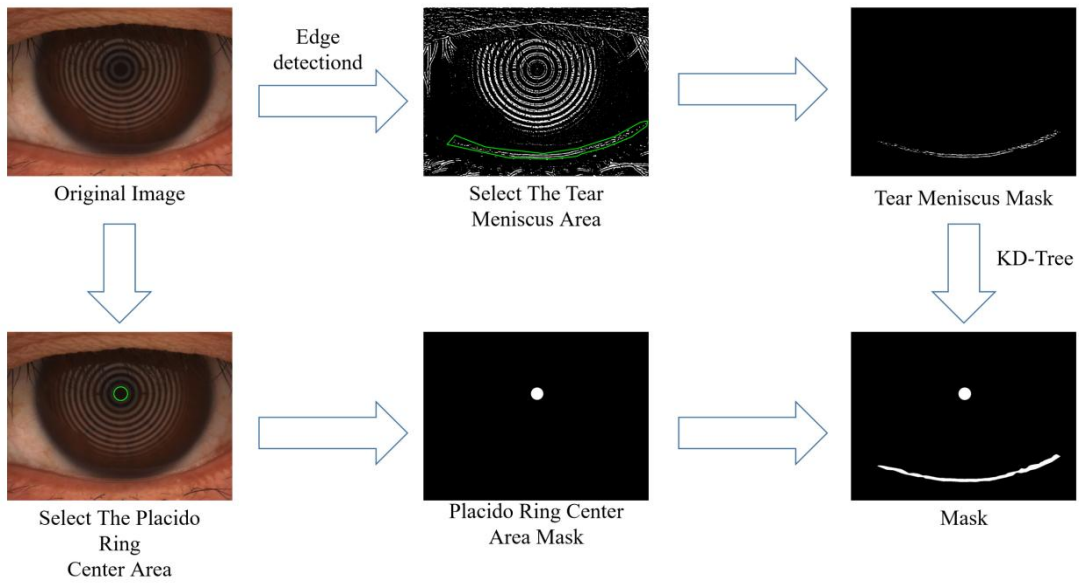

**Fig. 1.** Flowchart of edge detection-based mask generation. Edge detection aims to identify and highlight the boundaries of the tear meniscus. This is achieved by appropriately selecting the tear meniscus region and applying our algorithm for restoration. The repaired tear meniscus is then merged with the pupil area to create a unified mask label.

# 3. Model

## 3.1 Quality assessment model

Before calculating TMH with the algorithms, eliminating some images of inferior quality was necessary. In this study, images that are clear, without eyelashes obstructing the pupil and tear meniscus areas, and with complete and clear pupil and tear meniscus areas are defined as good-quality images. Conversely, images that are overly bright or dark, blurry, with closed eyes, or where the pupil or tear meniscus areas are severely obstructed by eyelashes, eyelids, or foreign objects are defined as poor-quality images. Following the quality assessment, 834 good-quality images and 166 poor-quality images were obtained. A total of 84 images were randomly selected from the 834 good-quality images and another 84 from poor-quality images. Both sets were used to train the image quality assessment model.

Data augmentation was performed on the premise of ensuring data balance. This included appropriate translation and rotation, as well as simulating real-life scenarios by adjusting image brightness and adding Gaussian noise to varying degrees. The augmented data was divided into a training set and a validation set at a ratio of 7:3. This study used four deep learning models for image quality assessments. The model with the highest accuracy was ultimately selected as the quality filtering model.

## 3.2 Edge detection operator

Edge detection, fundamental to many image processing tasks, preserves the original framework information of an image, thereby simplifying its analysis and processing. For example, Canny [16] proposed an operator focused on detecting brightness edges, which is a classic edge detection operator that has been widely studied and used. Ren et al. [17] improved contour detection accuracy by calculating Sparse Code Gradients (SCG). Furthermore, numerous deep learning-based edge detection methods have been developed, such as the Bi-Directional Cascade Network (BDCN) structure introduced by He et al. [18], which incorporates a Scale Enhancement Module (SEM) to merge multi-scale information for edge detection. Additionally, Lei et al. [19] proposed a Fully Convolutional Siamese Network (FCSN) for estimating joint object boundary flow in videos. We used edge detection and other preprocessing methods in this study to initially separate the tear meniscus area, thereby simplifying mask generation and addressing challenges such as difficulty in obtaining annotations and low image quality in medical image segmentation tasks.

First, the original image underwent sequential convolution with edge detection and filtering operators to highlight the tear meniscus against the background. The calculation formula is:

$$Out = Input \bullet EDO_{k_1} \bullet FO_{k_2},$$

where $Out$ represents the output image, $Input$ represents the input image, while $EDO_{k_1}$ and $FO_{k_2}$ represents the customized detection and filtering operators, respectively.

$$EDO_{k_1} = \begin{bmatrix} a_{11} & a_{12} & \cdots & a_{1k_1} \\ a_{21} & a_{22} & \cdots & a_{2k_1} \\ \vdots & \vdots & \ddots & \vdots \\ a_{k_1 1} & a_{k_1 2} & \cdots & a_{k_1 k_1} \end{bmatrix}, FO_{k_2} = \begin{bmatrix} b_{11} & b_{12} & \cdots & b_{1k_2} \\ b_{21} & b_{22} & \cdots & b_{2k_2} \\ \vdots & \vdots & \ddots & \vdots \\ b_{k_2 1} & b_{k_2 2} & \cdots & b_{k_2 k_2} \end{bmatrix},$$

$$a_{ij} = \begin{cases} -1, & i \neq \frac{k_1+1}{2} \text{ or } j \neq \frac{k_1+1}{2} \\ k_1^2 - 5, & i, j = \frac{k_1+1}{2} \end{cases}, b_{ij} = \begin{cases} 0, & i \neq \frac{k_2+1}{2} \\ \frac{1}{k_2}, & i = \frac{k_2+1}{2} \end{cases},$$

In this study, we set $k_1 = 13$, and $k_2 = 7$, with the rank sum values of the matrix $a_{ij}$, $b_{ij}$ being determined by experiments. These operators were customized specifically to highlight the tear meniscus area against the background while simultaneously balancing image brightness and minimizing noise.

Subsequently, a polygon tool was used to select the approximate tear meniscus area and remove other parts from the processed images. A KD-Tree-based connectivity algorithm [20] was used to efficiently organize and retrieve data points in multi-dimensional space. It traversed each point of the target, connecting it with the nearest $k$ points to effectively repair the tear meniscus. This approach enhanced the efficiency and quality of the generated tear meniscus area mask. Finally, the repaired tear meniscus was combined with the pupil into a single mask.

Experts B and C independently measured the TMH in 834 images. This measurement was repeated three times, with intervals between each session, to minimize the influence of previous measurements. The TMH data measured by both experts were combined. If no significant disagreement in the measurements was observed, the ground-truth of the TMH for that image was established by averaging the values. In cases of significant disagreement, consultation with a more experienced doctor was required to determine the ground-truth of the TMH.

## 3.3 image segmentation models

The 834 data points were divided into training, validation, and test sets in a 5:2:3 ratio, and data augmentation was performed. The images were cropped by 136 pixels on both the left and right sides to yield an image size of 3×1024×1024. Subsequently, the image size was adjusted to 3×512×512 for input into the convolutional neural network for training. This process aimed to identify the best model weights for generating the mask to predict the input image data. The workflow of the model is shown in Fig. 2.

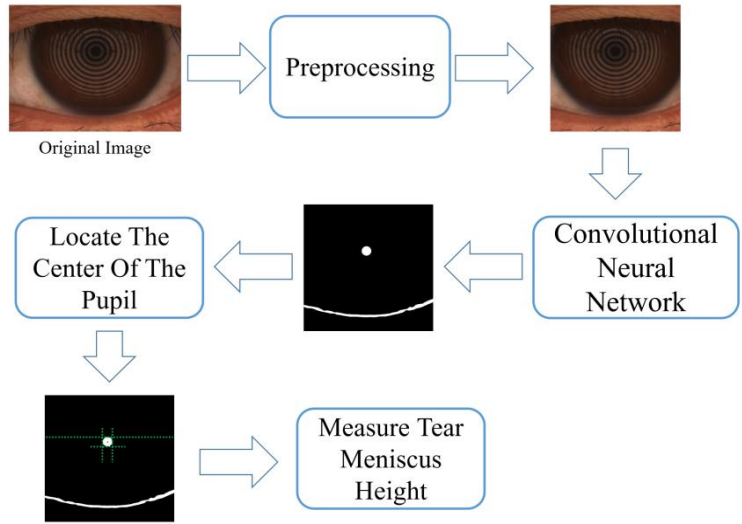

**Fig. 2.** Schematic diagram of the algorithm workflow. The original image is evenly cropped from left and right sides in a size of 1024×1024. After rescaling the image size to 512×512, a convolutional neural network is then employed to predict tear meniscus and pupil masks. Subsequently, the center of the pupil is localized, and the height of the tear meniscus is measured.

Several convolutional neural networks were trained in this study, with Unet [21] exhibiting the best performance. The Unet algorithm used in this study comprised nine convolutional blocks, four upsampling layers, and four downsampling layers. Each convolutional block included a 3×3 convolutional layer with a stride of 1, a batch normalization layer, a dropout layer, and an activation layer with LeakyReLU as the activation function, and the layers were stacked twice sequentially. The upsampling layers each consisted of a 1×1 convolutional layer with a stride of 1, followed by nearest-neighbor interpolation for upsampling. The downsampling layers each included a 3×3 convolutional layer with a stride of 2, a batch normalization layer, and an activation layer, which were stacked sequentially for downsampling. The final output was obtained through a 3×3 convolutional layer with a stride of 1 and a Sigmoid layer. Reflect was used as the padding mode in the convolutional and downsampling blocks, while nearest neighbor interpolation was used in the upsampling layers. Additionally, we experimented with incorporating deformable convolutional layers to replace the original convolutional layers and adding new cascading structures; however, these modifications did not yield significant improvements and substantially increased computational complexity. Ultimately, Unet was determined to be the most effective network architecture. The specific network structure of the Unet used in this study is shown in Fig. 3.

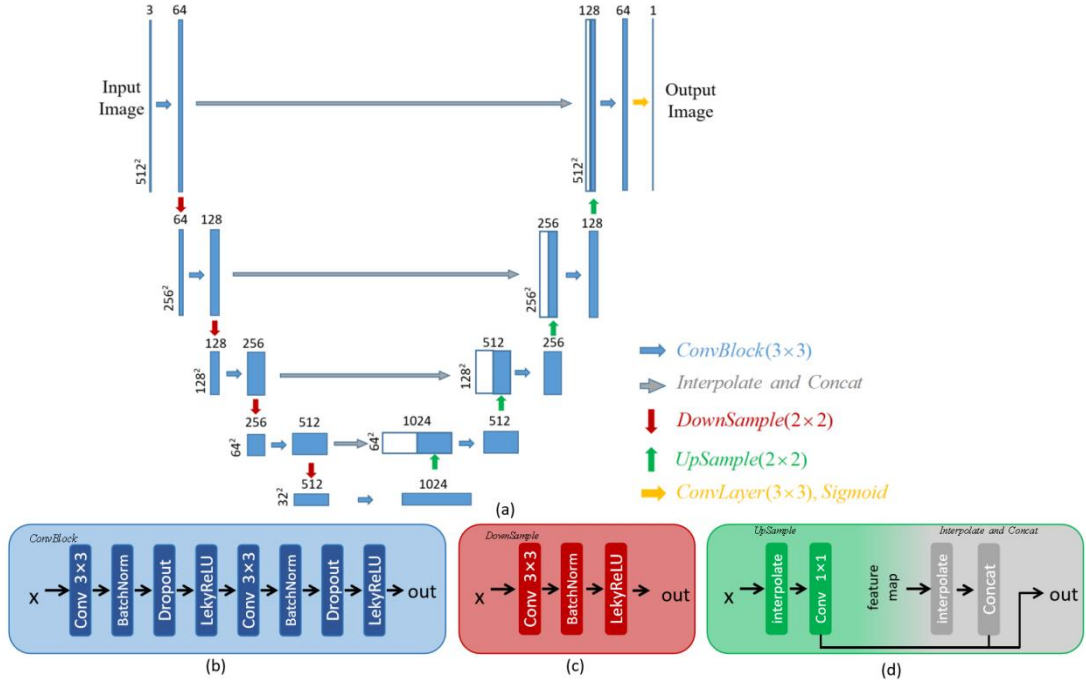

Fig. 3. (a) Schematic diagram of the network model structure, (b) ConvBlock, (c) DownSample, and (d) Interpolate and Concat layers

## 3.4 Loss function and optimizer

The loss function in this study is a combination of Binary Cross-Entropy Loss (BCELoss), Dice Loss (DiceLoss), and Matrix Norm Loss (MatrixNormLoss) functions. BCELoss aided in better fitting the model to the training data, DiceLoss enhanced segmentation and object control in images, and MatrixNormLoss regulated the scale and complexity of model parameters, thereby improving the model's stability and generalization ability. The formula of the loss function is as follows:

$$BCELoss = AVE(\sum_{i,j} -y_{ij} \cdot \log(y'_{ij}) - (1 - y_{ij}) \cdot \log(1 - y'_{ij})) \ ,$$

$$DiceLoss = 1 - \frac{2|Y \cap Y'|}{|Y| + |Y'|},$$

$$MatrixLoss = \|Y - Y'\|_2 = \sqrt{\lambda_{\max}(Y - Y')^T(Y - Y')},$$

$$LossFunction = 0.45 \cdot BCELoss + 0.45 \cdot DiceLoss + 0.1 \cdot MatrixLoss,$$

where $AVE(\cdot)$ represents the mean, $y'_{ij}$ represents the value of the model's predicted image at coordinate $(i, j)$, $y_{ij}$ represents the value of the annotated mask at coordinate $(i, j)$, $Y$ is the model's output prediction mask, $Y'$ is the ground-truth of the mask, $|Y|$ represents the number of $Y$ non-zero elements, $|Y'|$ represents the number of non-zero elements in $Y'$, and $|Y \cap Y'|$ represents the number of intersecting non-zero elements between $Y$ and $Y'$.

The model in this study adopted the Adam optimizer, leveraging its fast adaptive learning rate and momentum to facilitate rapid and stable training of deep neural networks. The specific parameters are as follows:

Learning Rate (lr) = 0.0001. Adam uses exponentially decaying moving averages to estimate the first-order moment (mean, $\alpha$) and the second-order moment (mean square, $\beta$) of the gradient, which are 0.9 and 0.999, respectively. To ensure numerical stability and avoid division by zero, $\varepsilon = 1e-8$ was added to the denominator.

## 3.5 Calculation of TMH

TMH was defined as the distance between the lower boundary of the pupil center and the upper and lower boundaries of the tear meniscus. Deng et al. [11] demonstrated that when the length, d, of the measurement section ranges from 0.5 mm to 4 mm, the TMH measurements are highly robust. In this study, a mask consisting of the pupil and tear meniscus areas was generated using Unet. This mask was then resized and binarized, and the highest point of the pupil area was identified. From this point, a 160×160 square area was created downwards that fully encompassed the pupil area. The algorithm automatically determined the center coordinates of the non-zero pixels within this square, corresponding to the coordinates of the pupil center. Coordinates of non-zero pixels at the same horizontal level as the pupil area were excluded from the entire mask, and the coordinates of the tear meniscus area remained. By selecting a measurement section length of d=0.5 mm and using the tear meniscus coordinates within this section, the TMH was calculated.

In clinical practice, various methods are used by doctors to measure TMH. This study, therefore, investigated two key questions: (1) whether significant differences are present in the results obtained by doctors using different TMH measurement methods, and (2) whether significant differences are present in the results obtained by algorithms using different measurement methods. The following three measurement methods are summarized in Fig. 4.

Firstly, let $TMRS$ represent the set of coordinates in the tear meniscus area,

$$TMRS = \{(x_i, y_{ij}) \mid i=1,2,\cdots,n \quad j=1,2,\cdots,m_i\}, \quad PC = x_k,$$

where $n$ represents the number of pixels spanned by the tear meniscus area, $y_{ij}$ represents the y-coordinate of the $j$ th point at x-coordinate $x_i$, $m_i$ represents the number of points at x-coordinate $x_i$, and $PC$ represents the x-coordinate of the pupil center.

Method 1: A measurement section of length d based on the pupil center coordinates was taken in the tear meniscus area. The difference between the highest and lowest y-coordinates within the measurement section was calculated and averaged to determine TMH. This is the method commonly used in the literature to measure TMH [11,12], and the formula is:

$$TMH = ave(\sum_{i=k-d}^{k+d} \max(y_{ij}) - \min(y_{ij})),$$

where $ave(\cdot)$ represents the average, $\max(\cdot)$ represents the maximum value, and $\min(\cdot)$ represents the minimum value.

Method 2: A measurement section of length d based on the pupil center coordinates was taken in the tear meniscus area. The upper and lower boundaries of the tear meniscus were fitted. For each point on the upper boundary, the point on the lower boundary with the closest tangent slope within a certain range was identified, and the average distance between these points was calculated. The formula is:

$$y_{up} = f_1(x), \; y_{low} = f_2(x), \; (x_k - 5 < x < x_k + 5),$$

$$\hat{x}_i = \arg\min(f_2'(\hat{x}) - f_1'(x_i)), \; (x_k - 5 < \hat{x} < x_k + 5),$$

$$TMH = ave(\sum_{i=k-5}^{k+5} \sqrt{(x_i - \hat{x}_i)^2 + (f_1(x_i) - f_2(\hat{x}_i))^2}),$$

where $y_{up}$ and $y_{low}$ represent the fitted curves of the upper and lower boundaries of the tear meniscus within the measurement section.

Method 3: A measurement section of length d based on the pupil center coordinates was taken in the tear meniscus area. The lengths of the upper and lower boundary curves in this area were calculated using numerical integration, and TMH was determined using the area of the measurement section. The formula is:

$$TMH = \frac{2S}{L_{up} + L_{low}},$$

where $S$ represents the area of the tear meniscus in the measurement section, and $L_{up}$ and $L_{low}$ represent the lengths of the given upper and lower boundary curves of the tear meniscus area. In subsequent analysis, we will compare the results between the first two manual measurement methods and also compare the measurement results of the above three measurement methods (i.e., Methods 1, 2, and 3), aiming to explore the differences between these various methods.

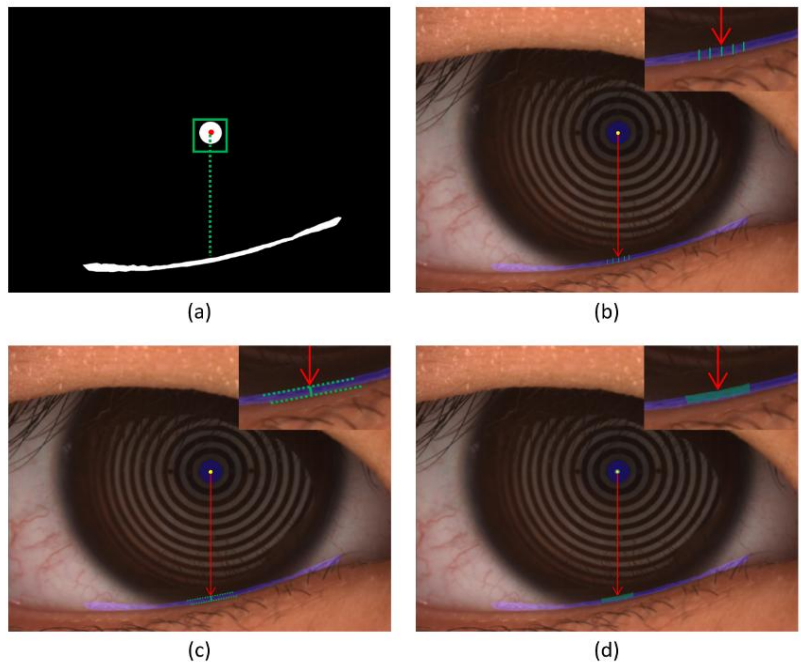

**Fig. 4.** Schematic diagram of various TMH measurement methods. (a) Schematic diagram of pupil center positioning and the tear meniscus, (b) Method 1 for TMH calculation, (c) Method 2 for TMH calculation, and (d) Method 3 of TMH calculation. TMH conversion coefficient: c=0.011575, unit: mm per pixel (mm/p).

# 4. Results

## 4.1 Data Annotation Quality Assessment

Here we will conduct statistical analysis on annotations by different experts to assess the reliability and differences. The quartile description and differential analysis of the TMH and the horizontal coordinates of the pupil center measured by the three experts are summarized in Table 1. The MIoU of masks generated by the three experts was compared, and the results are shown in Table 2.

**Table 1.**

Statistical description and differential analysis of the TMH and the horizontal coordinate of the pupil center measured by three experts

|   | Descriptive Statistics | | Pearson correlation | | Spearman rank correlation | | ICC($c_1$) |
|---|---|---|---|---|---|---|---|
|   | upper quartile | lower quartile | r-value | p-value | r-value | p-value |  |
| A | 15.12 | 21.15 | 0.976 |  | 0.975 |  | 0.975 |
| B | 14.23 | 21.67 | 0.956 | < 0.001 | 0.942 | < 0.001 | 0.955 |
| C | 14.00 | 21.33 | 0.978 |  | 0.970 |  | 0.974 |
| U |  |  | 0.964 |  | 0.955 |  | 0.995 |

(a) Comparison of TMH

|   | Descriptive Statistics | | Pearson correlation coefficient | | Spearman correlation coefficient | | ICC($c_1$) |
|---|---|---|---|---|---|---|---|
|   | upper quartile | lower quartile | r-value | p-value | r-value | p-value |  |
| A | 628.83 | 725.44 |  |  |  |  |  |
| B | 629.08 | 725.58 | 0.999 | <0.001 | 0.999 | <0.001 | 0.999 |
| C | 628.83 | 725.00 |  |  |  |  |  |

(b) Comparison of the horizontal coordinates of the pupil center

A, B, and C represent different experts. Statistical descriptions were based on the average quartile (calculated as the mean of the upper quartile values from three sets of measurements by the corresponding expert and the mean of the lower quartile values). The Pearson correlation test, Spearman correlation test, and ICC were used to evaluate differences. All data were calculated and assessed in pixels to avoid excessively small values.

**Table 2.**

Comparison of MIoU of masks generated by the three experts

|   | A | B | C |
|---|---|---|---|
|   | MIoU | MIoU | MIoU |
| A | 0.8133 | 0.7727 | 0.7518 |
| B | 0.7727 | 0.8191 | 0.7780 |
| C | 0.7518 | 0.7780 | 0.8702 |

A, B, and C represent different experts. Data in the table represent MIoU between masks generated by two experts.

Analysis of the data in Tables 1 and 2 reveals high intra-group and inter-group correlations for annotations made by the three experts, including the horizontal coordinates of the pupil center, TMH, and masks. Notably, the intra-group correlation for the masks generated by Expert C is the highest among the three groups and requires the least time.

In summary, the edge detection approach for mask generation demonstrates high accuracy. It can replace the pixel-by-pixel method of mask generation and significantly reduce the influence of subjective factors from the annotators. In addition to enhancing robustness, the method substantially decreased the workload of mask generation, thereby increasing work efficiency.

## 4.2 Networks performance

In the collected images, challenges such as out-of-focus, dim or overly bright lighting, machine mode settings, and subjects blinking complicated the identification of the tear meniscus and pupil. This quality assessment model will serve as a reminder to doctors of the need to retake images, thereby improving work efficiency and the precision of subsequent measurements. Further optimization of the quality assessment model can be achieved by expanding the dataset to improve the effectiveness in assessing the quality of acquired images, and the results are shown in Table 3. Among the image quality assessment models, Inceptionv3 was the best-performing model, attaining an accuracy of 98.224%. Assessment results of the output masks from different network models are summarized in Table 4. In this study, the Unet network structure demonstrated the best performance.

**Table 3.**

Accuracy of the quality assessment models

| Model | Accuracy |
| --- | --- |
| Densenet121 | 97.862% |
| Resnet50 | 97.230% |
| Resnet152 | 96.662% |
| Inceptionv3 | 98.224% |

**Table 4.**

Performance assessment of output masks from different network models

| | MIoU | recall | precision | F1-Score |
| --- | --- | --- | --- | --- |
| **Unet[21]** | **0.9362** | **0.9261** | **0.9423** | **0.9326** |
| ResUnet[22] | 0.9185 | 0.9141 | 0.9134 | 0.9120 |
| DeepLab3+fcnresnet101[23] | 0.8817 | 0.8509 | 0.8890 | 0.8674 |
| DeepLab3+fcnresnet50 | 0.8826 | 0.8472 | 0.8954 | 0.8684 |
| FcnResnet50[24] | 0.9307 | 0.9245 | 0.9310 | 0.9262 |
| FcnResnet101 | 0.8621 | 0.8368 | 0.8533 | 0.8418 |

## 4.3 Reliability analysis of TMH

In this study, 200 images were randomly selected, and two experts measured TMH using Methods 1 and 2, respectively to compare the differences between these methods. The results indicate that the measurement differences between the methods are not significant. See Table 5. In addition, TMH was measured in 250 images from the test set using the three methods mentioned above. The ACC for each method was calculated. The results indicate that the measurement differences between the methods are not significant. See Table 6. Comparative analysis revealed that the differences in TMH calculated by the three methods are not significant. An appropriate method can be selected based on the actual situation for calculation.

**Table 5.**

Comparison of differences in TMH measurements between Method 1 and Method 2

| | Spearman correlation coefficient | | ICC | | ACC |
| --- | --- | --- | --- | --- | --- |
| | p-value | r-value | $c_1$ | $c_2$ | |
| A | 0.9914 | <0.001 | 0.9938 | 0.9952 | 1.00 |

|   | B | 0.9846 | <0.001 | 0.9895 | 0.9947 | 1.00 |

**Table 6.**

Accuracy of Unet combined with three TMH calculation methods

|   | ACC |
| --- | --- |
| Method1 | 0.9480 |
| Method2 | 0.9360 |
| Method3 | 0.8080 |

All measurements in this study underwent the Shapiro–Wilk test, with a p-value < 0.001 indicating non-normal distribution. As previously mentioned, the Unet model exhibited the best performance; therefore, we utilized its mask output for TMH measurement and assessment. As shown in Table 7, the results indicate that our algorithm (Ours) achieved a 94.80% accuracy (237/250) in calculating TMH, compared to 96.95% accuracy (191/197) by Expert A. Linear regression and Bland-Altman plots comparing our calculated TMH with the ground-truth and TMH measured by Expert A against the ground-truth are shown in Fig. 5. The model measured TMH with 97.60% (244/250) of points within the 95% CI, with the fitting curve equation being y=0.982x-0.862. Expert A's measurements had 93.91%(185/197) of points within the 95% CI, with the fitting curve equation being y=0.976x-0.257. These results indicate a significant correlation between the model output TMH and the ground-truth, and high consistency with the TMH measured by Expert A, and the accuracy values were highly similar.

**Table 7.**

Statistical comparison of correlation between TMH and ground-truth measured by our algorithm (Ours) and Expert A

|   | Pearson correlation test | | Spearman rank test | | Linear regression | | | ICC | | ACC |
| --- | --- | --- | --- | --- | --- | --- | --- | --- | --- | --- |
|   | r-value | p-value | r-value | p-value | k | b | $r^2$ | $c_1$ | $c_2$ |   |
| Ours | 0.9912 | <0.001 | 0.9883 | <0.001 | 0.982 | -0.862 | 0.961 | 0.981 | 0.990 | 0.9480 |
| A | 0.9869 | <0.001 | 0.9861 | <0.001 | 0.976 | -0.257 | 0.961 | 0.981 | 0.990 | 0.9695 |

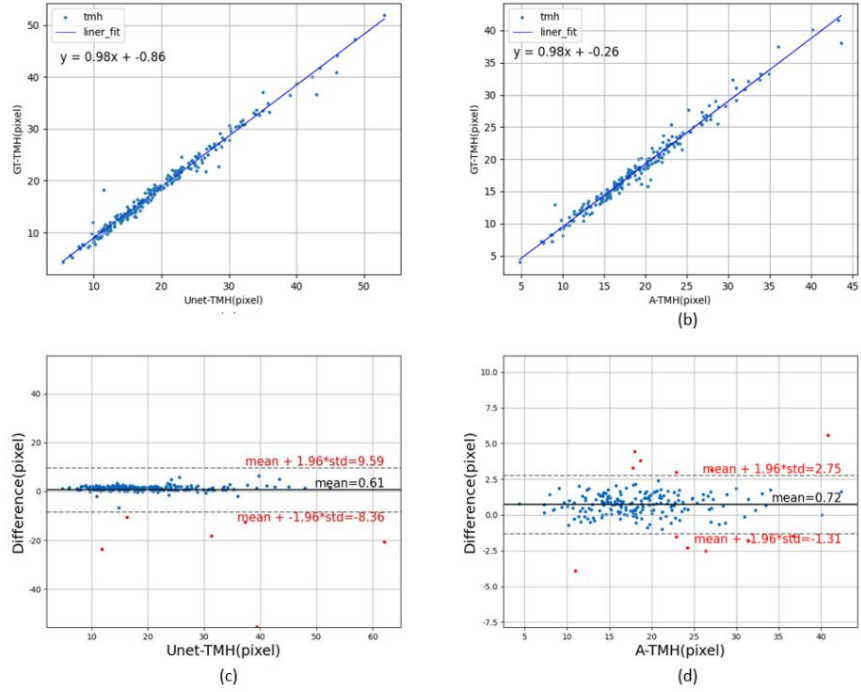

**Fig 5.** (a) Linear regression of TMH calculated using the algorithm versus ground-truth, (b) linear regression of TMH measured by Expert A versus ground-truth, (c) Bland-Altman plot of TMH calculated by the algorithm versus ground-truth, with 97.60% (244/250) points within the 95% CI, and (d) Bland-Altman plot of TMH measured by Expert A versus ground-truth, with 93.91% (185/197)of points within the 95% CI.

# 5. Discussion

DED is a prevalent eye disease that significantly impacts the visual function of patients, and its detrimental effects should not be underestimated [25]. Research indicates that TMH is a critical parameter in assessing the tear meniscus and is essential in diagnosing DED. However, a standardized method for TMH measurements still lacking. Existing methods are primarily manual or semi-automatic, which are not only laborious and time-consuming but also suffer from poor repeatability, potentially resulting in inaccurate diagnoses [26]. Even with deep learning models that automatically calculate TMH, the training annotations require manual, pixel-by-pixel input by operators using annotation software [11,12]. This process is labor- and time-intensive, with inherent subjective factors that may lead to inaccurate results.

The algorithm developed in this study uses masks based on edge detection methods. By processing images with edge detection, which maintains their original positional information and structure, the algorithm preliminarily separates the tear meniscus from the background. This approach significantly enhances the efficiency and accuracy of annotation and reduces subjective biases in the annotation process.

The algorithm incorporates the Inceptionv3 network architecture for quality assessment and classification and the Unet network architecture for segmenting images that meet quality standards. Additionally, the algorithm can calculate TMH by simultaneously segmenting the pupil and tear meniscus areas. Comprehensive evaluations suggest that the TMH calculated by the algorithm is highly consistent with that determined by experts. The quality assessment model had an accuracy of 98.224%, the overall MIoU for the segmentation model was 0.9362, and the accuracy of TMH calculation was 0.9480. When compared with the ground-truth, the Pearson correlation coefficient was 0.9912 (p<0.001), and the Spearman correlation coefficient was 0.9883 (p<0.001), with the linear fitting curve expression being $y = 0.982x - 0.862$, $r^2 = 0.961$.

During the study of TMH calculations, we observed that although the tear meniscus shapes at both ends and the center of some eyes are relatively consistent, certain images exhibited excessively wide areas on either side of the tear meniscus. This variability may be attributed to differences caused by other ocular diseases or factors. García-Resúa et al. suggested investigating the correlation between various tear meniscus shapes and TMH through subjective classification [27]. Improvements based on the segmentation of the tear meniscus area could potentially transform this subjective classification into a more objective approach to investigate any association between the tear meniscus shape and other ocular conditions. The performance of the quality assessment model performance is currently limited by the small dataset used for training. The model can only be applied to eliminate images of extremely poor quality and can hardly differentiate images from patients with conjunctivochalas. Expanding the dataset may enhance the performance of the classification model. In medical image segmentation tasks, the use of edge detection operators could address challenges such as the difficulty in acquiring mask annotations and the issue of low annotation quality. Additionally, TMH is only one important indicator in diagnosing DED. Integration of TMH with other relevant indicators could lead to the development of a more comprehensive multimodal AI system for DED diagnosis, aiding doctors in both diagnosis and treatment of the condition.

# 6. Conclusion

In this study, we proposed a deep learning method based on edge detection for the automatic assessment of image quality, as well as for the automatic segmentation and measurement of TMH in the tear meniscus and pupil areas. This method enhances the efficiency and objectivity of the image annotation process, automates and refines the calculation of TMH, and integrates various TMH calculation logics that can be selected based on actual conditions. The results indicate that the TMH measurements obtained using our method exhibited high consistency with those obtained through manual measurement. As new data are continuously incorporated, weight parameters in the original network will be adjusted, and the model's generalizability and robustness are expected to improve. This improvement will significantly enhance the efficiency of clinical screening for DED.

**Author contributions**

KW: algorithm design, statistical evaluation and drafting the manuscript. KX: data collection, data analysis. XC: data analysis and statistical evaluation. CH: mathematical model and revising the manuscript. JZ: data analysis and statistical evaluation. DK: study suggestions and revising the manuscript. QD: study design and revising the manuscript. SH: study design and revising the manuscript.

**Acknowledgements:**

This research was funded by the grants from Zhejiang Normal University (Grant Nos. YS304222929, YS304222977, ZZ323205020522016004), and the National Natural Science Foundation of China (Grant Nos. 12301676, 12090020 and 12090025), and the Zhejiang Provincial Medical and Health Science Technology Program, Health Commission of Zhejiang Province (Grant No: 2022PY074).

**Data Availability Statement:**

The data presented in this study are available on request from the corresponding authors.

**Declaration of Competing Interest:**

The authors declare that they have no conflict of interest.

# References

[1] J.P. Craig, K.K. Nichols, E.K. Akpek, B. Caffery, H.S. Dua, C.K. Joo, Z. Liu, J. D. Nelson, J.J. Nichols, K. Tsubota, et al., TFOS DEWS II definition and classification report, Ocul. Surf. 15 (2017) 276–283.

[2] A.J.Bron, C.S. de Paiva, S.K.Chauhan, S. Bonini, E.E. Gabison, S. Jain, E. Knop, M. Markoulli, Y. Ogawa, V. Perez, Y. Uchino, N. Yokoi, D. Zoukhri, D.A. Sullivan., TFOS DEWS II pathophysiology report, Ocul Surf. 15(2017)438-510.

[3] F. Stapleton, M. Alves, V.Y. Bunya, I. Jalbert, K. Lekhanont, F. Malet, K.S. Na, D. Schaumberg, M. Uchino, J. Vehof, E. Viso, S. Vitale, L. Jones, TFOS DEWS II Epidemiology Report, Ocul Surf. 15(2017)334-365.

[4] A. Uchida, M. Uchino, E. Goto, E. Hosaka, Y. Kasuya, K. Fukagawa, M. Dogru, Y. Ogawa, K. Tsubota, Noninvasive interference tear meniscometry in dry eye patients with Sjögren syndrome, Am J Ophthalmol. 144(2007)232-237.

[5] Y. Yuan, J. Wang, Q. Chen, A. Tao, M. Shen, M. Abou Shousha, Reduced tear meniscus dynamics in dry eye patients with aqueous tear deficiency, Am. J. Ophthalmol. 149 (2010) 932–938.

[6] L. Tian, J. Qu, X. Sun, et al., Repeatability and reproducibility of noninvasive keratograph 5M measurements in patients with dry eye disease, J. Ophthalmol. 2016 (2016).

[7] H. Stegmann,v.a. Dos Santos, A. Messner, A. Unterhuber, D. Schmidl, G. Garhöfer, L. Schmetterer, R.M. Werkmeister, Automatic assessment of tear film and tear meniscus parameters in healthy subjects using ultrahigh-resolution optical coherence tomography, Biomed Opt Express. 10(2019)2744-2756.

[8] H. Stegmann, R.M. Werkmeister, M. Pfister, G. Garhöfer, L. Schmetterer, V.A. Dos Santos, Deep learning segmentation for optical coherence tomography measurements of the lower tear meniscus, Biomed Opt Express. 11(2020)1539-1554.

[9] R. Arita, K. Yabusaki, T. Hirono, T. Yamauchi, T. Ichihashi, S. Fukuoka, N. Morishige, Automated Measurement of Tear Meniscus Height with the Kowa DR-1α Tear Interferometer in Both Healthy Subjects and Dry Eye Patients, Invest Ophthalmol Vis Sci. 60(2019)2092-2101.

[10] J. Yang, X. Zhu, Y. Liu, X. Jiang, J. Fu, X. Ren, K. Li, W. Qiu, X. Li, J. Yao, TMIS: a new image-based software application for the measurement of tear meniscus height, Acta Ophthalmol. 97 (2019) e973–e980.

[11] X. Deng, L. Tian, Z. Liu, Y. Zhou, Y. Jie, A deep learning approach for the quantification of lower tear meniscus height, Biomed Signal Process Control. 68(2021), 102655.

[12] C. Wan, R. Hua, P. Guo, P. Lin, J. Wang, W. Yang, X. Hong, Measurement method of tear meniscus height based on deep learning, Front Med (Lausanne).14(2023);10, 1126754.

[13] T.K. Koo, M.Y. Li, A Guideline of Selecting and Reporting Intraclass Correlation Coefficients for Reliability Research, Journal of Chiropractic Medicine. 15(2016)155-163.

[14] C. Spearman, The proof and measurement of association between two things, The American Journal of Psychology, 15(1904) 72–101.

[15] J.M. Bland, D.G. Altman. Statistical methods for assessing agreement between two methods of clinical measurement, Lancet. 1(1986)307-10.

[16] J. Canny, A Computational Approach to Edge Detection, IEEE Transactions on Pattern Analysis and Machine Intelligence.PAMI-8(1986) 679-698.

[17] X.F. Ren, B. Liefeng, Discriminatively Trained Sparse Code Gradients for Contour Detection, Neural Information Processing Systems.1(2012) 584-592.

[18] J. He, S. Zhang, M. Yang, Y. Shan, T. Huang, BDCN: Bi-Directional Cascade Network for Perceptual Edge Detection, IEEE Trans Pattern Anal Mach Intell. 44(2022):100-113.

[19] P. Lei, F. Li, S. Todorovic, Boundary Flow: A Siamese Network that Predicts Boundary Motion Without Training on Motion, IEEE/CVF Conference on Computer Vision and Pattern Recognition. (2018) 3282-3290.

[20] A. Moore.An introductory tutorial on kd-trees, IEEE Colloquium on Quantum Computing: Theory, Applications

&Implications,IET, (1991).

[21] O. Ronneberger, P. Fischer, T. Brox, U-net: convolutional networks for biomedical image segmentation, Int. Conf. Med. Image Comput. Comput. Interv. (2015) 234–241.

[22] K. He, X. Zhang, S. Ren and J. Sun, Deep Residual Learning for Image Recognition, IEEE Conference on Computer Vision and Pattern Recognition (CVPR). (2016) 770-778.

[23] Chen, LC., Zhu, Y., Papandreou, G., Schroff, F., Adam, H., Encoder-Decoder with Atrous Separable Convolution for Semantic Image Segmentation.Computer Vision11211(2018)pringer, Cham. https://doi.org/10.1007/978-3-030-01234-2_49.

[24] J. Long, E. Shelhamer and T. Darrell, Fully convolutional networks for semantic segmentation,IEEE Conference on Computer Vision and Pattern Recognition (CVPR). (2015), 3431-3440.

[25] The epidemiology of dry eye disease: report of the Epidemiology Subcommittee of the International Dry Eye WorkShop (2007). Ocul Surf. 2007 Apr;5(2):93-107. doi: 10.1016/s1542-0124(12)70082-4. PMID: 17508117.

[26] J C Mainstone, A S Bruce, T R Golding, Tear meniscus measurement in the diagnosis of dry eye, Current Eye Research, 15(1996) 653-661.

[27] C. García-Resúa, J. Santodomingo-Rubido, M. Lira, M.J. Giraldez, E.Y. Vilar, Clinical assessment of the lower tear meniscus height, Ophthalmic Physiol Opt.29(2009)487-496.